# The power–law TST reaction rate coefficient with tunneling correction

Yin Cangtao, Zhou Yanjun and Du Jiulin[*]

*Department of Physics, School of Science, Tianjin University, Tianjin 300072, China*

**Abstract:** We study the TST reaction rate for the systems with power-law distributions. We derive the expressions of the reaction rate coefficient with tunneling correction, which strongly depends on the power-law parameter. The numerical results show that a small deviation from one in the parameter can result in a significant change in the rate coefficient, but only cause a small change in the tunneling correction. Thus the tunneling correction is not sensitive to the power-law distributions. As an application example, we take $H+H_2$ reaction to calculate the power-law reaction rate coefficient with the tunneling correction, the results of which with the parameter slightly different from one are in good agreement with all the experimental studies in temperature range $2\times10^2 \sim 10^3$ K.

**Keywords**：Reaction rate coefficient, power-law distribution, tunneling correction, nonequilibrium system

## 1. Introduction

Transition state theory (TST) has made it possible to obtain quick estimates for the reaction rates of a broad variety of processes in natural sciences and engineering technology, and thus it has became a cornerstone and a core of the reaction rate theory. In previous work [1], we derived a generalized TST reaction rate formula for an elementary reaction, but without consideration of the tunneling effect,

$$k_{\nu-\text{cl}} = \omega_f \left[ 2 + \left(\frac{k_B T}{h\omega_f}\right)^{\nu-1} (Z_\Delta^{\neq})^{\nu-1} - \sum_i Z_i^{\nu-1} - (\nu-1)\frac{\Delta\varepsilon}{k_B T} \right]^{\frac{1}{\nu-1}}, \quad (1)$$

where the subscript cl denotes the classical result without the tunneling effect, $\omega_f$ is the reaction coordinate frequency (i.e. decomposition frequency of transition state, then $1/\omega_f$ is mean lifetime of transition state), $h$ is Planck constant, $k_B$ is Boltzmann constant, $T$ is temperature, $Z_\Delta^{\neq}$ is the transition state partition function which has removed the vibration partition function, $Z_i$ is the partition function of the $i$th reactant molecule, $\Delta\varepsilon$ is difference of basic energies between the transition state and reactants, and the parameter $\nu\neq1$ represents the power–law distributions in nonextensive statistical mechanics (NSM) [2, 3]. As expected, in the limit $\nu\rightarrow1$ Eq. (1) can be well reduced to the standard TST reaction rate formula in Boltzmann–Gibbs (BG) statistics [4],

$$k_{1-\text{cl}} = k_{\text{TST-cl}} = \frac{k_B T}{h} \frac{Z_A^{\neq}}{\prod_i Z_i} \exp\left(-\frac{\Delta\varepsilon}{k_B T}\right). \tag{2}$$

---

[*] Email address:

Usually, because the protons can tunnel through the activation energy barrier which separates reactants from products, the effective height of the barrier would decrease and the probability of the proton transfer reaction would increase. This is the tunneling effect, which can be described by drawing the wave function of the protons near the barrier. Actually, the proton tunneling becomes important only at the low temperature when most of the reactants are trapped on the left of the barrier [5]. Therefore, if we study the reaction rate of the proton (or hydrogen) reactions at low temperature, we should take the tunneling effect into account.

This work is to study the tunneling effect with power-law distributions and add the tunneling correction to the power-law TST reaction rate formula. Therefore, in Section 2, we study the tunneling correction for the power-law $\nu$-distribution and the corresponding TST reaction rate coefficient. In Section 3, we numerically analyze the power-law rate coefficient with the tunneling correction. In Section 4, as an application example, we calculate the rate coefficients of $H+H_2$ reaction and compare with the experimental studies. Finally in Section 5, we give the conclusion and discussion.

**2. The power–law TST rate coefficient with tunneling correction**

The tunneling effect is about the passage of the particles through the energy barrier that is greater than their kinetic energy. When the particles lack sufficient kinetic energy to go over the barrier through the minimum energy pathway, by means of the tunneling effect the chemical reactions can also occur. If the tunneling effect is not considered, the classical trajectory method underestimates the rate coefficient when the energy of the system is close to the classical energy threshold of the reactions [6]. For a BG distribution, if the tunnel correction $\kappa$ is added, the reaction rate formula is usually revised as

$$k_{BG} = \kappa k_{TST-cl} \tag{3}$$

. Therefore, for the power-law $\nu$-distribution, Eq.(3) should be generalized to that

$$k_\nu = \kappa_\nu k_{\nu-cl}, \tag{4}$$

with the tunnel correction $\kappa_\nu$. In the previous work [1], we derived the power-law TST rate coefficient $k_{\nu-cl}$ for the elementary reaction. Now we will study the tunnel correction $\kappa_\nu$ for the power-law $\nu$-distribution.

The potential energy barrier, which approaches closely the classical reaction path of the atom transfer reactions and has an analytical solution for the transmission probability, is Eckart barrier [7],

$$V(x) = \frac{A \exp(x/b)}{1+\exp(x/b)} + \frac{B \exp(x/b)}{(1+\exp(x/b))^2}. \tag{5}$$

When $A = 0$, the barrier becomes symmetrical and has its maximum at $x = 0$ and $\Delta\varepsilon = B/4$.

According to quantum mechanics, the transmission probability of the particle with energy $\varepsilon$ to go through Eckart barrier [8] is that,

$$G(\varepsilon) = \frac{\cosh(2\alpha\zeta^{1/2})-1}{\cosh(2\alpha\zeta^{1/2})-\cosh(\sqrt{4\alpha^2-\pi^2})}, \tag{6}$$

where the parameters are $\zeta = \varepsilon/\Delta\varepsilon$, and $\alpha = 2\pi\Delta\varepsilon/h\omega_f$.

In a real chemical reaction system, the reactant molecules with different energies follow a distribution at temperature $T$, and thus the total transmission probability is the integral of $G(\varepsilon)$ over all the energies. In the conventional theory, the reactant molecules and the transition state are both assumed to be a BG distribution. Based on this assumption, the tunneling correction is taken [6–8] as the exponential form,

$$\kappa = \frac{1}{k_B T} \exp\left(\frac{\Delta\varepsilon}{k_B T}\right) \int_0^\infty d\varepsilon\, G(\varepsilon) \exp\left(-\frac{\varepsilon}{k_B T}\right). \tag{7}$$

However, complex systems are generally far away from equilibrium. In the physical, chemical, biological and engineering technical processes, there have been a lot of theoretical works and experimental studies to reveal the power-law distributions (see [9] and the references therein), such as single-molecule conformational dynamics [10, 11], reaction-diffusion processes [12], chemical reactions [13], gene expressions [14] and those listed in NSM [15]. In these processes, the reaction rates may be energy-dependent (and/or time-dependent [16, 17]) power–law forms [18, 19], which thus are beyond the scope governed by the conventional reaction rate formulae for a BG distribution. In these situations, the traditional reaction rate theories and the rate formulae have to be modified and generalized so that they are applicable. Recently, the research has made progress in the TST reaction rate theory [18], the collision theory reaction rate coefficients [20], the unimolecular reaction rate coefficients [21] and the barrierless reaction rate coefficient [22]. In addition, the mean first passage time [23] and the escape rate [24, 25] for the power-law distributions in both overdamped systems and low-to-intermediate damping were also studied. As we can imagine, the reaction rate theory for power-law distributions is a complicated and exciting field in exploring the understanding of nonequilibrium reaction rate theory.

Chemical reaction involves a large number of particles following a statistical probability distribution. Similarly to reference [1], we can replace the BG distribution in BG statistics by the power-law distribution in NSM [15]. On the basis of NSM, the energy distribution function can be written [20] by power-law $\nu$-distribution,

$$\frac{dN(\varepsilon)}{N_0} = \frac{\nu}{k_B T}\left[1-(\nu-1)\frac{\varepsilon}{k_B T}\right]^{\frac{1}{\nu-1}} d\varepsilon, \qquad (8)$$

where $N(\varepsilon)$ is the number of particles with energy ranging from $\varepsilon$ to $\varepsilon+d\varepsilon$, $N_0$ is the total particle number.

If $J_0$ is the total flux of the particles striking the left hand side of the barrier, and $J$ is the quantum mechanical rate of the particles appearing on the right hand side of the barrier, then [7, 8] we have that

$$J = J_0 \frac{\nu}{k_B T} \int_0^\infty d\varepsilon \, G(\varepsilon) \left[ 1 - (\nu-1)\frac{\varepsilon}{k_B T} \right]^{\frac{1}{\nu-1}}. \tag{9}$$

In the classical mechanics framework, $G(\varepsilon) = 0$ for $\varepsilon < \Delta\varepsilon$ and $G(\varepsilon) = 1$ for $\varepsilon > \Delta\varepsilon$, the classical rate $J_{cl}$ is

$$J_{cl} = J_0 \frac{\nu}{k_B T} \int_{\Delta\varepsilon}^\infty d\varepsilon \left[ 1 - (\nu-1)\frac{\varepsilon}{k_B T} \right]^{\frac{1}{\nu-1}}$$

$$= J_0 \left[ 1 - (\nu-1)\frac{\Delta\varepsilon}{k_B T} \right]^{\frac{\nu}{\nu-1}}. \tag{10}$$

As usual, the tunneling correction $\kappa_\nu(T)$ is also defined by the ratio of the quantum mechanical rate to the classical rate. Using Eq. (9) and Eq.(10), we can find the tunneling correction,

$$\kappa_\nu = \frac{J}{J_{cl}} = \frac{\nu}{k_B T} \left[ 1 - (\nu-1)\frac{\Delta\varepsilon}{k_B T} \right]^{\frac{-\nu}{\nu-1}} \int_0^\infty d\varepsilon \, G(\varepsilon) \left[ 1 - (\nu-1)\frac{\varepsilon}{k_B T} \right]^{\frac{1}{\nu-1}} \tag{11}$$

If we take the limit $\nu \to 1$, Eq.(11) is perfectly reduced to the standard form Eq. (7) in the conventional theory.

Finally, combining Eq.(1), Eq.(4), Eq.(6) and Eq.(11), we derive the power-law TST reaction rate formula with the tunneling correction,

$$k_\nu = \frac{\nu \omega_f}{k_B T} \left[ 2 + \left(\frac{k_B T}{\hbar \omega_f}\right)^{\nu-1} (Z_\Delta^{\neq})^{\nu-1} - \sum_i Z_i^{\nu-1} - (\nu-1)\frac{\Delta\varepsilon}{k_B T} \right]^{\frac{1}{\nu-1}} \left[ 1 - (\nu-1)\frac{\Delta\varepsilon}{k_B T} \right]^{\frac{-\nu}{\nu-1}}$$

$$\times \int_0^\infty d\varepsilon \frac{\cosh(2\alpha\zeta^{1/2}) - 1}{\cosh(2\alpha\zeta^{1/2}) - \cosh(\sqrt{4\alpha^2 - \pi^2})} \left[ 1 - (\nu-1)\frac{\varepsilon}{k_B T} \right]^{\frac{1}{\nu-1}}. \tag{12}$$

As expected, in the limit $\nu \to 1$ Eq.(12) recovers the standard form of the TST reaction rate formula with tunneling correction in BG statistics [7, 8],

$$k_{BG} = \frac{1}{h} \frac{Z_\Delta^{\neq}}{\prod_i Z_i} \int_0^\infty d\varepsilon \frac{\cosh(2\alpha\zeta^{1/2}) - 1}{\cosh(2\alpha\zeta^{1/2}) - \cosh(\sqrt{4\alpha^2 - \pi^2})} \exp\left(\frac{-\varepsilon}{k_B T}\right). \tag{13}$$

**3. Numerical analysis of the power–law rate formula with tunneling correction**

In order to show the dependence of the power-law rate coefficient $k_\nu$, Eq.(12),

and the tunneling correction $\kappa_\nu$, Eq.(11), on the $\nu$-parameter and temperature $T$, we numerically analyzed these equations, the results of which were shown in Figs.1-4. In the numerical analysis, the fixed values were taken from the data in the reaction, $H+H_2 \rightarrow H_2+H$, which as an example of the application will be discussed in section 4. For example [6], we have chosen respectively $\Delta\varepsilon = 38.2$ kJ mol$^{-1}$, $T = 300$ K and $\omega_f = 1493$ cm$^{-1}$ as the fixed values of the energy difference, the temperature and the reaction coordinate frequency. The parameters $\zeta$ and $\alpha$ in Eq.(12) are calculated by using $\zeta = \varepsilon/\Delta\varepsilon$ and $\alpha = 2\pi\Delta\varepsilon/h\omega_f$ in Eq.(6).

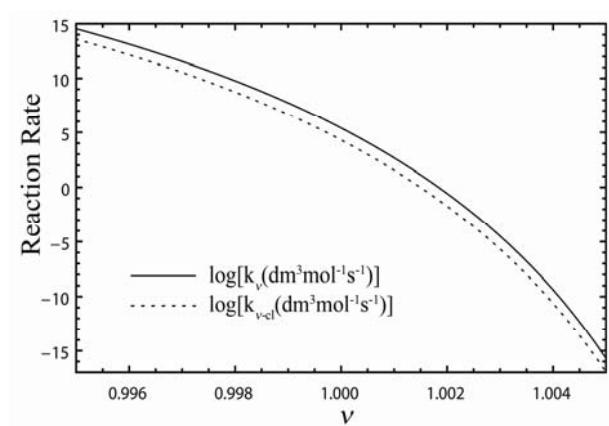

Fig. 1. Dependence of the reaction rate coefficients on the $\nu$-parameter

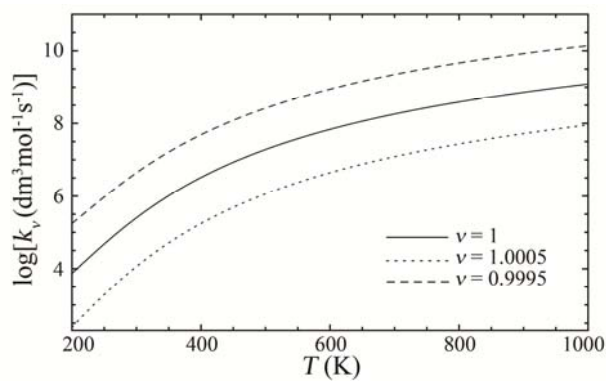

Fig. 2. Dependence of the rate coefficient $k_\nu$ on temperature $T$
for three values of the $\nu$-parameter

In Fig.1, we illustrated two lines, respectively, to show the dependences on the $\nu$-parameter of the rate coefficient $k_\nu$ in Eq.(12) with the tunneling correction and the rate coefficient $k_{\nu-cl}$ in Eq.(1) without the tunneling correction, where $\Delta\varepsilon =38.2$ kJ mol$^{-1}$, $T = 300$ K and $\omega_f = 1493$ cm$^{-1}$. The reaction rate axis is plotted on logarithmic scale. The $\nu$–axis is ranging 0.995~1.005, implying the state not very far away from BG distribution. Fig.1 showed very strong dependences of the rate coefficients on the $\nu$-parameter: when the $\nu$-parameter deviated from 1 very small ($\nu=0.995~1.005$), the rate coefficient changed in a very large range, $k_\nu \approx 10^{15}$~$10^{-15}$ dm$^3$mol$^{-1}$s$^{-1}$. Thus a small deviation from BG distribution would result in a very significant variation in the reaction rate. We also find that the tunneling correction exists, but it actually is not large and the rate coefficient $k_\nu$ with the tunneling correction is slightly larger than that $k_{\nu-cl}$ without the tunneling correction.

In Fig.2 we illustrated the dependence of the rate coefficient $k_\nu$ in Eq.(12) with the tunneling correction on the temperature for three values of the $\nu$-parameter, where $\Delta\varepsilon=38.2$ kJ mol$^{-1}$ and $\omega_f = 1493$ cm$^{-1}$. The $k_\nu$–axis is plotted on logarithmic scale, and the $T$-axis is taken as 200~1000 K, ranging the temperature of all the experimental studies of the chemical reactions in NIST chemical kinetics database at http://kinetics.nist.gov/kinetics. Fig.2 showed a significant effects of $\nu\neq 1$ on $k_\nu$, where the curve of $\nu=1$ corresponds to the rate coefficient $k_{BG}$ in Eq.(13) for the conventional TST based on BG statistics. The property of the rate coefficient about the temperature is basically the same as $k_{\nu-cl}$ in Eq.(1) without the tunneling correction [1].

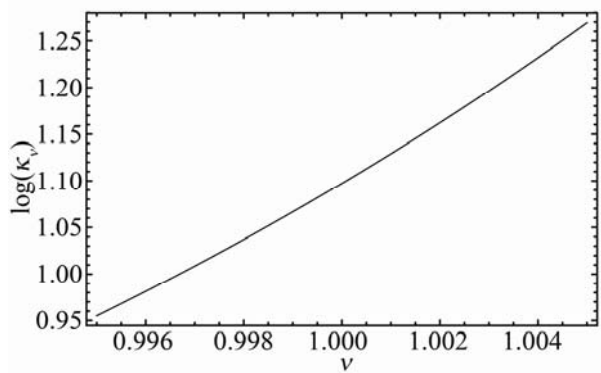

Fig. 3. Dependence of the tunneling correction $\kappa_\nu$ on the $\nu$-parameter

In Fig.3, we illustrated the dependence of the power-law tunneling correction $\kappa_\nu$

in Eq.(11) on the $\nu$-parameter. The temperature is $T$=300 K. It is shown that when the $\nu$-parameter is taken $\nu$=0.995~1.005, the tunneling correction varies only $\kappa_\nu \approx 10^{0.95}$~$10^{1.25}$, which tell us that the tunneling correction is not sensitivity to the $\nu$-parameter as compared with the rate coefficients in Fig.1.

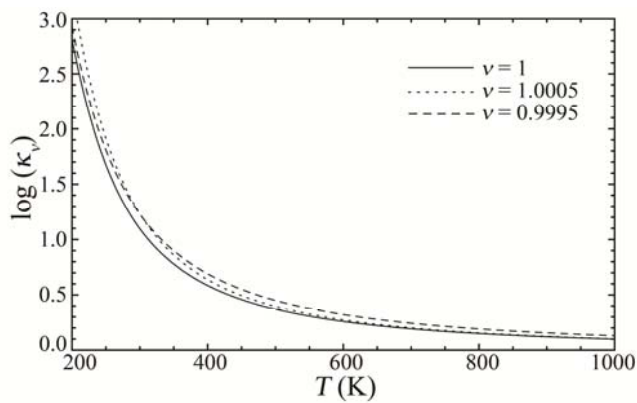

Fig. 4. Dependence of the tunneling correction $\kappa_\nu$ on temperature $T$
for three values of the $\nu$-parameter

In Fig.4, we illustrated the dependence of the power-law tunneling correction $\kappa_\nu$ in Eq.(11) on the temperature $T$ for three values of the $\nu$-parameter, where the $T$-axis was taken as 200~1000 K, ranging the temperature in all the experimental studies, and the curve $\nu$=1 corresponds to the tunneling correction in Eq.(7) for the conventional TST in BG statistics. As usual, the tunneling correction becomes significant only under low temperature. Fig.4 showed the effect of the parameter $\nu \neq 1$ on the tunneling correction $\kappa_\nu$ and this effect is minor for small deviations, $\nu$=1.0005 and $\nu$=0.9995.

**4. Application to H+H$_2 \rightarrow$ H$_2$+H reaction**

As an application of the power-law reaction rate coefficient, Eq.(12), to the chemical reactions in the systems with the power-law $\nu$-distribution, we now take H+H$_2$ reaction to calculate the rate coefficient. The elementary reaction process of this reaction [6] is that

$$\text{H+H}_2 \xrightarrow{k_\nu} \text{H}_2\text{+H} . \qquad (14)$$

The activated complex (i.e., the transition state) is $\text{H}\cdots\text{H}\cdots\text{H}$. In Table 1, we listed some properties of the reactants and the activated complex. There are two degenerate bending: $\omega_1$ is at the transition state and $\omega_2$ is the symmetric stretching. The anti-symmetric stretching $\omega_3$ becomes the reaction coordinate and has an imaginary value, marked with $i$ in the table. $I$ is the moment of inertia. The symmetry number is $\sigma$ for calculations of the rotations.

Table 1. Properties of the reactants and activated complex in H+H$_2$ reaction [6]

| Parameters | H…H…H | H | H$_2$ |
|---|---|---|---|
| $r$ (H–H) / nm | 0.09287 | | 0.0741 |
| $\omega_1$ / cm$^{-1}$ | 899 | | 4400 |
| $\omega_2$ / cm$^{-1}$ | 2067 | | |
| $\omega_3$ / cm$^{-1}$ | 1493$i$ | | |
| $\Delta\varepsilon$ / kJ·mol$^{-1}$ | 38.2 | | |
| $m$ /10$^{-27}$ kg | 5.022 | 1.674 | 3.348 |
| $I$ / (10$^{-48}$ kg·m$^2$) | 28.876 | | 4.596 |
| $\sigma$ | 2 | | 2 |

Table 2. The experimental and theoretical values of the rate coefficients of H+H$_2$ reaction

| $T$(K) | $k_{BG}$ | $\kappa$ | $k_{exp}$ | $\delta$ | $k_\nu$ | $\kappa_\nu$ | $\nu$ |
|---|---|---|---|---|---|---|---|
| 200 | 7.453×10$^3$ | 653.8 | 5.809×10$^3$ | 28.3% | 5.809×10$^3$ | 660.0 | 1.0000396 |
| 225 | 1.925×10$^4$ | 139.1 | 2.395×10$^4$ | 19.6% | 2.395×10$^4$ | 138.2 | 0.9999641 |
| 250 | 4.798×10$^4$ | 47.18 | 7.438×10$^4$ | 35.5% | 7.438×10$^4$ | 46.74 | 0.9999264 |
| 300 | 2.523×10$^5$ | 12.51 | 4.071×10$^5$ | 38.1% | 4.071×10$^5$ | 12.43 | 0.9999169 |
| 500 | 1.912×10$^7$ | 2.344 | 1.220×10$^7$ | 56.7% | 1.220×10$^7$ | 2.347 | 1.0000831 |
| 800 | 3.883×10$^8$ | 1.412 | 8.255×10$^7$ | 370% | 8.255×10$^7$ | 1.413 | 1.0002967 |
| 1000 | 1.221×10$^9$ | 1.257 | 1.561×10$^8$ | 682% | 1.561×10$^8$ | 1.258 | 1.0003994 |

The rate coefficients of this reaction in all the experimental studies were taken from NIST chemical kinetics database at http://kinetics.nist.gov/kinetics, and they were satisfied by the two-parameter fit, namely

$$k_{exp} = 3.32 \times 10^{-12} \exp(-21200/RT) \text{cm}^3\text{molecule}^{-1}\text{s}^{-1} \tag{15}$$

with RMSD=5.0. All the experimental studies were carried out in the temperature range: $2\times10^2 \sim 10^3$ K. The steps to get Eq. (15) in the database are that, firstly, we input H2 and H in the left hand-side of the arrow and then input H and H2 in the right hand-side of the arrow, secondly, we click the "submit" button and then check all the 2nd order experiments, and finally, we click the "create plot" button and can get the experimental data.

In Table 2, we listed the experimental and theoretical values of the rate coefficients of H+H$_2$ reaction of the temperature $2\times 10^2 \sim 10^3$ K, where $k_{BG}$ was calculated using Eq.(13), $k_{exp}$ was obtained using Eq. (15), and $k_\nu$ was calculated using Eq.(12). The quantity $\delta$ is defined as the relative error of $k_{BG}$ to $k_{exp}$ by $\delta = |k_{BG} - k_{exp}| / k_{exp}$. The units of the rate coefficients are dm$^3$mol$^{-1}$s$^{-1}$. We also listed the values of the tunneling correction $\kappa$ in Eq.(7) and the power-law tunneling correction $\kappa_\nu$ in Eq.(11) for this reaction.

Table 2 shows that there are significant relative errors of $k_{BG}$ to $k_{exp}$, but $k_\nu$ with the $\nu$-parameter slightly different from 1 can be in good agreement with all the experimental studies. There are only minor differences arising due to the $\nu$-parameter between $\kappa_\nu$ and $\kappa$.

Table 2 also shows that the fitted value of the $\nu$-parameter varies with the change of the temperature. The variation in the $\nu$-parameter is a result of the fact that the $\nu$-parameter may not only depend on the intermolecular interactions but also on the temperature, mirroring the differences between the experimental studies at different temperature and their environment. Small deviation from one in the $\nu$-parameter (thus from a BG distribution) can result in significant change in the rate coefficient, but only produce minor change in the tunneling correction. Therefore, the tunneling correction is not sensitive to the power-law $\nu$-distribution.

## 5. Conclusion

In conclusion, we have studied the power-law reaction rate coefficient with the tunneling correction for the systems with the power-law $\nu$-distribution. We have derived the power-law tunneling correction Eq.(11) and the power-law rate coefficient Eq.(12) with the tunneling correction, which all depend on the $\nu$-parameter.

The numerical results have illustrated the properties of the power-law tunneling correction $\kappa_\nu$ and the corresponding rate coefficient $k_\nu$ with the tunneling correction. Based on the new formulae, Eq.(11) and Eq.(12), and the old formulae, Eq.(7) and Eq.(1), we find that small deviation from 1 in the $\nu$-parameter (thus from a BG distribution) can result in significant change in the rate coefficient, but can not produce drastic change in the tunneling correction. The power-law rate coefficient with the tunneling correction is slightly larger than that without the tunneling correction. Therefore, the tunneling correction is not sensitive to the power-law $\nu$-distribution.

We have taken H+$H_2$→$H_2$+H reaction as an application example to calculate the power-law reaction rate coefficient $k_\nu$ based on Eq.(12) with the tunneling correction. For this reaction, we show that the power-law rate coefficient $k_\nu$ with the $\nu$-parameter slightly different from 1 can be in good agreement with all the experimental studies in temperature range $2\times10^2\sim10^3$ K.

**Acknowledgment**

This work is supported by the National Natural Science Foundation of China under Grant No. 11175128 and also by the Higher School Specialized Research Fund for Doctoral Program under Grant No. 20110032110058.